# Time-series analysis of fissure-fed multi-vent activity: a snapshot from the July 2014 eruption of Etna volcano (Italy)

Spina[1,2*], L., J. Taddeucci[3], A. Cannata[2,4], M. Sciotto[4], E. Del Bello[3], P. Scarlato[3], U. Kueppers[1], D. Andronico[4], E. Privitera[4], T. Ricci[3], J. Pena-Fernandez[5], J. Sesterhenn[5], D.B. Dingwell[1]

[1]*Ludwig-Maximilians-Universität München (LMU), Theresienstrasse 41, 80333 Munich, Germany*
[2] *Dipartimento di Fisica e Geologia, Università degli Studi di Perugia, Piazza dell'Università 1, 06123, Perugia, Italy*
[3] *Istituto Nazionale di Geofisica e Vulcanologia, Via Di Vigna Murata 605, 00143, Roma, Italy*
[4] *Istituto Nazionale di Geofisica e Vulcanologia, Sezione di Catania, Osservatorio Etneo, Piazza Roma 2, 95125, Catania, Italy*
[5] *Technische Universität Berlin, Institut für Strömungsmechanik und Technische Akustik, Müller-Breslau Straße 15, 10587 Berlin, Germany*

*Corresponding author:
Dr. Laura Spina,
email:laura.spina@unipg.it; phone: +39 075 585 2607
address: Dipartimento di Fisica e Geologia, Università degli Studi di Perugia, Piazza dell'Università 1, 06123, Perugia, Italy

**Abstract:**

On 5 July 2014, an eruptive fissure opened on the eastern flank of Etna volcano (Italy) at ~3.000 m a.s.l. Strombolian activity and lava effusion occurred simultaneously at two neighbouring vents. In the following weeks, eruptive activity led to the build-up of two cones, tens of meters high, here named Crater N and Crater S. To characterize the short-term (days) dynamics of this multi-vent system, we performed a multi-parametric investigation by means of a dense instrumental network. The experimental setup, deployed on July 15-16$^{th}$ at ca. 300 m from the eruption site, comprised two broadband seismometers and three microphones as well as high speed video and thermal cameras. Thermal analyses enabled us to characterize the style of eruptive activity at each vent. In particular, explosive activity at Crater N featured higher thermal amplitudes and a lower explosion frequency than at Crater S. Several episodes of switching between puffing and Strombolian activity were noted at Crater S through both visual observation and thermal data; oppositely, Crater N exhibited a quasi-periodic activity. The quantification of the eruptive style of each vent enabled us to infer the geometry of the eruptive system: a branched conduit, prone to rapid changes of gas flux accommodated at the most inclined conduit (i.e. Crater S). Accordingly, we were able to correctly interpret acoustic data and thereby extend the characterization of this two-vent system.

# 1. Introduction

Knowing the structure of a volcano's plumbing system is essential for realistically assessing volcanic hazards. The shallow feeding system not only represents the transient path for magmatic fluids towards the surface, but it is also the theatre of fundamental processes influencing the style of activity, i.e. volatile exsolution and magma crystallization. The complexity of the plumbing system also impacts on the distribution of active vents; as a consequence, volcanic activity often takes place through a complex of multiple vents. In the case of mafic magmas, multiple vents with shifting activity characterize both century-old systems at persistently active volcanoes, like Stromboli (Italy; e.g. Ripepe and Marchetti, 2002) and Yasur (Vanuatu; e.g. Spina et al., 2016), and shorter-lived flank or eccentric eruptions, like Pu'u'O'o' (Hawai'i; e.g. Garces et al., 2003) and Etna in 2002-2003 (Italy; e.g. Andronico et al., 2009). The potential interplay among different vents has implications for the interpretation of geophysical signals and thereby for monitoring as well as for visualising the subsurface volcanic structure (Johnson et al., 2005). Hence, the temporal evolution of closely-spaced vents along an eruptive fissure at Etna has been investigated through infrasound signals, yielding constraints on the geometry of the plumbing system during the May-September 2008 activity (Cannata et al., 2011a). The investigation of conduit processes is strongly enhanced by a multi-parametric approach that integrates infrasound with seismic, thermal, gas flux and video measurements (Johnson et al., 2004 and references therein).

Here, we performed a multi-parametric experiment at Etna in order to investigate the dynamic processes at the two-vent system active during the initial phase of the sub-terminal eruption (i.e. an eruption located very close to the summit craters and normally fed directly by the same feeding conduits; e.g. Badalamenti et al., 2004) that occurred in summer 2014. This eruption took place within a phase of elevated activity from 2011 to 2015, with tens of lava-fountaining episodes, lava effusion, and Strombolian activity from some of Etna's summit craters (Behncke et al., 2014; Andronico et al., 2015).

On July 5th 2014, an eruptive fissure (hereafter EF) opened at ~3000 m elevation on the lower eastern flank of the North East Crater (NEC; **Fig. 1c**), 700 m far from the New South East Crater (NSEC, **Fig. 1c**; De Beni et al., 2015). The eruptive activity was initially characterized by both the effusion of lava flows and Strombolian explosions occurring at two vents. After three weeks, the explosive activity had led to the formation of two cones a few tens of meters high (INGV-OE Internal Report, 2014a), hereafter called Crater N and Crater S (**Fig. 1a; Fig. 2c, d**). On July 23th the two cones had merged and the activity at the two vents subsequently occurred within a single horseshoe-shaped complex, which fed a lava flow at its base (INGV-OE Internal Report, 2014b). On July 25th, a new vent, featuring Strombolian activity and lava flows, opened at a short distance upslope at 3090 m of elevation (INGV-OE Internal Report, 2014b). From the first days of August on, the intensity of eruptive activity diminished and came to an end on August 10th (INGV-OE Internal Report, 2014c). From August 8th -16th, Strombolian activity was observed at the NSEC (INGV-OE Internal Report, 2014c). The sub-terminal eruption of July-August 2014 has been linked to the NSEC, inferred amongst others from the system of dry fractures developed between the EF and NSEC, and propagating from south to northwest (De Beni et al., 2015). A sketch of the eruption history is shown in **Fig. 2a** together with the envelope of seismic Root Mean Square (RMS) amplitude (**Figs.1b**, **2b**). Increases in RMS, related to intensification of volcanic tremor amplitude, accompanied the opening of the July 5th and July 25th eruptive vents and the apex of explosive episodes of early August at the EF and NSEC (**Fig. 2b**).

Our multi-parametric experiment was realized on July 15-16th, when the two vents were feeding a rather stable Strombolian activity, with the ejection of gas, and lapilli- to bomb-sized pyroclasts to a maximum height of 100-200 meters above the crater rim in the near absence of ash. At that time, Crater N and Crater S were roughly 20 and 15 m high over the surroundings, respectively (**Fig. 2c**). The goal of this study is to characterize the short-term (1-2 days) style of activity of the two vents simultaneously active along EF and the interplay between their dynamics, by coupling thermal videos with infrasound measurements. We demonstrate the advantage of such integrated approach, which provides the ability to distinguish the contribution of each vent, for the interpretation of geophysical signals.

## 2. Experiment setup

On July 15-16th 2014, the following instruments were deployed ~300 m away from the eruptive vents (**Fig. 1a**): 2 broadband three-component seismometers (Trillium Nanometrics$^{TM}$), 2 microphones (GRAS 40AN), one high-speed camera (Optronis CR600, results not reported in this study), one thermal camera (FLIR SC655) and one drone (DJI Phantom 2). From 7:00 to 12:00 on July 16th (all times are in GMT), one seismometer and one microphone (i.e., Station 1 in **Fig. 1a**) acquired in synch with the thermal camera, recording discontinuously several seconds- to minutes-long traces. The thermal camera generally operated at 50 fps, with a higher resolution (100 fps) between 08:40 and 10:15 a.m. Three different lenses were used (**Fig. 3b**). All the other instruments (i.e., Station 2 in **Fig. 1a**) recorded continuously from ~12:00 on July 15th to ~13:00 on July 16th. All microphones have a sensitivity of 50 mV/Pa in the range 0.3–20,000 Hz.

The seismic and acoustic signals were also recorded by the permanent network of INGV, Sezione di Catania, Osservatorio Etneo. Seismic data from the ECNE station, located ~1 km away from the EF, were used to estimate the long-term temporal variation of the seismic RMS (**Figs. 1b, 2b**). Very high-level signals due to explosion air-waves rendered the seismic data from the temporary seismometers unusable.

## 3. Data analysis

In the following sections, the analyses performed on the data recorded during the experiment are described.

### 3.1 Thermal analysis

### 3.1.1 Characterization of the eruptive vents

Direct observation of the eruptive activity at the two craters matches video derived eruption features (**Fig. 3**). Visually, Crater N was more active, featuring continuous Strombolian activity with the ejection of abundant pyroclasts, whereas at Crater S weaker Strombolian activity alternated with periods of gas-dominated, pyroclast-poor puffing activity (**Fig. 3a**).

Quantitatively, we used thermal videos enclosing both vents to record the thermal amplitude above the two craters: first, we defined two rectangular control areas above the rim of each crater, set as close as possible above the vents and appropriately sized to capture at best the gas and pyroclasts emitted by the vent (e.g. **Fig. 2d,** the size of the control areas has been scaled to account for the different lenses). For each frame, we integrated the amplitude of the thermal signal within the control area and removed the background temperature, defined as the average pixel temperature of a background area with no pyroclasts. Finally, we divided this by the number of pixel in the area to obtain the average pixel temperature, i.e. mean brightness temperature. To measure each thermal pulse, we applied a static threshold at zero on the mean brightness amplitude series, previously centred and cleared by any possibly present underlying trend by removing the mean of the signal. Such values were visually proofed to maximize the detection results. The resulting thermal series feature a sequence of peaks, each corresponding to an individual explosion ejecting incandescent pyroclasts and/or gas within the control area (**Fig. 4**). About 4,100 and 5,100 explosive events were identified for Crater N and S, respectively, over almost three hours of recording during the morning of July $16^{th}$. Crater N features a thermal amplitude of explosive events centred on a quite constant mean value. On the contrary, Crater S displays a sharp increase in amplitude at 07:47, visually linked to the transition from puffing, ejecting mostly gas, to Strombolian activity ejecting abundant pyroclasts (**Fig. 3b**). Furthermore, a slight decrease in amplitude occurs at Crater S at around 10:45-11:10, when brief return to puffing activity was observed (**Fig. 3a**, **3b**).

To quantify the relative contribution of hot gas versus pyroclasts to the amplitude of the thermal peak, we designed a low-pass smoothing filter. Pyroclasts are spatially limited, whereas the gas phase has a more diffuse signal. For the filter, we defined a square filter mask of a size larger than the apparent size of the largest pyroclast. We then assigned the lowest value found in the mask to each pixel within the mask. This sets pyroclast-containing pixels to gas or background values. The mask is applied to the whole image. Noteworthy, the ratio of thermal amplitude without/with the smoothing filter is expected to increase in pyroclastic-rich eruptions. Accordingly, intervals of pyroclast-poor puffing activity (before 07:47 and between 10:45-11:00) are marked by low values of the ratio of the thermal amplitude without/with the smoothing filter (**Fig. 3c**).

Finally, we estimated the recurrence time (i.e. inter-arrival time) of explosive activity at each vent by measuring the time-lapse between the start of pairs of consecutive events (e.g. Strombolian activity or puffing). To highlight the general pattern of inter-arrival time and remove high frequency variations, we computed the average inter-arrival time for each data series in a moving window of 0.4 seconds with a step of 0.02 seconds (**Fig. 3d**). Please note that the effect of different lens magnifications has been properly accounted for, and played no relevant role on the investigated parameters (explosion amplitude and inter-arrival time). Also, no differences were observed when sampling with higher frame rate. Slightly shorter inter-arrival times are observed during puffing activity at Crater S, whereas activity at Crater N was steady.

Crater S displays a bimodal distribution of the amplitude of the detected events and of the inter-arrival time, deriving from the presence of two distinct classes of events (i.e. puffing and Strombolian activity, **Fig. 5d**). At Crater N, the amplitude distribution is essentially unimodal (**Fig. 5c**). The median values of the distribution of inter-arrival times correspond to 2.3 s for Crater N, and 1.6 s for Crater S. The latter displays a relatively broader distribution of peak values (1-2.1s), coherent with the frequent changes in the eruptive activity observed at this vent. In contrast, the remarkably narrow peak in the distribution of inter-arrival times of Crater N is indicative of a high level of stability of the ongoing activity.

### 3.1.2 Thermal evidence of related dynamics at the two vents

The investigation of the relationship between the dynamics of the two eruptive vents (Crater N and S) is advantaged by cross-correlating the timelines of inter-arrival times detected at the two vents as well as cross correlating the timelines of their thermal amplitude.

However, it can be difficult to interpret the statistical significance of the simple cross correlation coefficients, whose value depends also on parameters such as the number of signal samples. Hence, we applied the randomised cross correlation method, that allows evaluating the statistical significance of the retrieved cross correlation functions (e.g. Martini et al., 2009; Zuccarello et al., 2013). With this aim, the correlation between two time series is calculated several times, randomizing at each run the time of one series. In order to calculate mean $\hat{c}$ and standard deviation $\sigma_c$ of these randomized correlations, one has to bear in mind that the correlation coefficients (as they are a normalized quantity) do not follow a Gaussian distribution (Zuccarello et al., 2013). Thus, Fisher's z transform is used to transform the calculated correlation coefficients into z, a new variable following a Gaussian distribution (e.g. Saccorotti and Del Pezzo, 2000).

In our case, the randomised cross correlation method was applied to the time series of both inter-arrival times and amplitudes of the detected events. We calculated the correlation 10,000 times, randomizing at each run the times of the Crater N series. The whole dataset of inter-arrival times and amplitudes was taken into account as unique time series. The results show a clear positive correlation between the two vents, suggested by both the inter-arrival time and the amplitude series. In particular, it is evident how the cross correlation values between the non-randomised times series fall above the $\hat{c} \pm 3\sigma_c$ band (dashed black lines in **Fig. 6b, 6d**) suggesting a statistically significant positive relationship between both the inter-arrival time and the amplitude series.

### 3.2 Infrasonic analysis

The integration of thermal and infrasonic signals is well known to provide information on the dynamics of explosive activity at active volcanoes (e.g. Marchetti et al., 2009). Here, the investigation of infrasonic signals is additionally motivated by the possibility of extending our range of temporal observation to the time in which Station 2 (**Fig. 1a**), equipped with an acoustic sensor recording continuously, was active (12:00 on July $15^{th}$ to ~13:00 on July $16^{th}$).

The infrasonic signal was dominated by infrasonic events with simple N-shaped waveforms, characteristic of Strombolian activity (e.g. Dalton et al., 2010; **Fig. 7a**). To investigate the spectral characteristics of the infrasonic signals, we calculated a spectrum by Fast Fourier Transform (FFT) using 5.12 s-long windows with an overlap of 50%. All spectra within windows of 15 minutes were normalised, averaged, and plotted as spectrogram. The highest spectral peaks are in the 4-8 Hz band (**Fig. 7c**). They alternate with few short events characterized by a lower frequency content (1–3 Hz), which correspond with low amplitude infrasonic events and lower signal-to-noise ratio (**Fig. 7b**).

The infrasonic events accompanying the explosions were detected by the percentile-based method (Cannata et al., 2011b). The infrasonic signal was filtered in the band 4-8 Hz, i.e. the band of the highest spectral peaks of acoustic signature. The root mean square (RMS) envelope of the infrasonic signal was calculated on 0.15 s-long windows sliding by 0.01 s. The threshold was calculated as the $10^{th}$ percentile of the entire RMS time series, multiplied by 6. These parameters were chosen after having tested many different value combinations and having visually checked the detection results on selected signal windows. A total number of 77,000 events were detected.

Then, inter-arrival times of the infrasound events were calculated as the time interval between two subsequent events. A histogram of the number of infrasound event pairs versus the inter-arrival time values displays a non-symmetric bimodal distribution with a higher peak centred at 0.7 s and a lower one at 2.2 s (**Fig. 8a**). The same distribution analysis was performed taking into account only the events with peak-to-peak amplitudes higher than the $50^{th}$ percentile calculated on all the peak-to-peak amplitudes. The more energetic events show a unimodal distribution pattern centred at 2.2 s (**Fig. 8b**). To evaluate the time evolution of the inter-arrival times, the histograms were calculated within 15 min-long windows, normalised by the maximum value, gathered in a single matrix and visualised showing time in the x-axis, inter-arrival times in the y-axis and the number of infrasound event pairs with the colour scale. Such normalization maximizes our capability to picture the temporal evolution of the dominant class in the frequency distribution of inter-arrival times. The first hours of the investigated period (up to 19:00 on July 15$^{th}$) were mainly characterised by explosions with inter-arrival times of 0.7 s (**Fig. 8c, d**) . Successively, until about 05:00 on July 16$^{th}$, the range of the most frequent inter-arrival times broadened at 0.5-2.6 s. In a brief time-interval around 01:30 on July 16$^{th}$ the distribution became unimodal with a clear peak at 2.2 s (**Fig. 8e**) . After about 05:00 on Juy 16$^{th}$ and up to the end of the investigated period, most of the explosions showed again an inter-arrival time of 0.7 s, except for the short time interval at ~10:45, lasting for about 30 minutes and characterised again by a broader inter-arrival times range (0.5-2.6 s).

Finally, to investigate the recurrence pattern of infrasound events and explosions, we evaluated the coefficient of variation (i.e. COV), computed by the standard deviation calculated on the inter-arrival times divided by their mean value. A COV value equal to 0 implies perfect periodicity, COV close to 0 quasi-periodicity, COV = 1 Poissonian recurrence, i.e. unpredictability, and COV > 1 temporal clustering (e.g. Cannata et al., 2013). COV was calculated on 15 min-long time windows taking into account both all the explosions and the explosions with peak-to-peak amplitude higher than the 50$^{th}$ percentile (**Figs. 8f,g**). In the former case, the COV analysis shows steady values of 0.4-0.6, suggesting a behaviour in between quasi-periodic and Poissonian. In the latter case, lower COV values (0.2-0.3) can be observed in the interval 16:30 on July 15$^{th}$ – 07:30 on July 16$^{th}$, suggesting a quasi-periodic behaviour of the more energetic explosions.

### 4. Discussion

The July-August 2014 eruption at Etna volcano took place along a short eruptive fissure that, on July 5$^{th}$, had started as a two-vent system on the upper eastern slope of the volcano. The relatively easy access as well as regular and persistent activity at the two vents offered a good chance to investigate the short-time dynamics operating in the shallow plumbing system of a fissure-fed multi-vent explosive-effusive volcanic system. The role played by concomitant lava effusion on the overall mass flow rate is not investigated in this work.

### 4.1 Evidences of conduit geometry from thermal investigation

Thermal videos represent a powerful tool to characterize the style of eruptive activity at each vent. The integration of the amplitude of thermal images to obtain representative waveforms of the thermal style of the eruption has been widely used for investigating the eruptive dynamics (e.g. Ripepe et al., 2002; Johnson et al., 2005; Sahetapy-Engel et al., 2008). The mean brightness amplitude, measured from thermal images, results from the superimposition of several contributions, such as temperature, size and amount of pyroclasts/gas within the control area, their emissivity and the source-receiver atmospheric effects (e.g. Marchetti et al., 2009). Variations in the brightness temperature are related to the size and amount of ash/pyroclasts in the control area and can be used as a proxy for magnitude, in terms of mass loading, temperature and optical density of the explosions (Sahetapy-Engel et al., 2008).

Assuming a similar ejection temperature for the gas and magma at the two vents, the observed differences in their thermal series indicate that Strombolian explosions at Crater N featured an overall higher magnitude (larger mass of ejecta) and longer and more constant recurrence time than at Crater S (**Fig. 3b, d**). The latter repeatedly switched between gas-rich puffing and relatively more bomb-rich Strombolian activity, and featured an overall lower intensity and higher occurrence rate of the explosions (**Fig. 3**). In the past, different styles of explosive activity at closely-spaced vents have often been observed. For example, during the 2002-2003 flank eruption, simultaneous Strombolian explosions, degassing and puffing activity took place at adjacent vents along the N-fissure, lasting from hours to few days, being inferred to be related to the same flow conditions over most of their conduit length (Andronico et al., 2009). The joined thermal-infrasonic investigation set up during the 2014 activity may greatly improve the understanding of the dynamics producing fissure-fed multi-vent activity.

The above-mentioned differences in the style of eruptive activity pinpoint fundamental constraints on the geometry of the feeding system. The significant correlation between the activities at the two vents in terms of both amplitudes and inter-arrival times (**Fig. 6**) suggests a common source at depth, likely in the form of a unique conduit splitting in branches. Previous studies have shown the effect of pipe bifurcation on fluid dynamics of an ascending gas-liquid mixture, focusing on phase redistribution between the two-branched arms (e.g. Marti and Shoham, 1997). The fluid partition at T-junctions in a two-phase system is governed by several variables such as the geometry of the branches, the flow pattern, the inclination of the branched arm and the flow rates (e.g. Penmatcha et al., 1996). The angle and diameter of the branch play an important role, as the gas phase tends to upraise preferentially through the more inclined conduit (e.g. Penmatcha et al., 1996) while reduced branch diameter enhances liquid drag (e.g. Baker et al., 2007). For slug flow approaching a vertical junction, the liquid phase entering in the branched arm with the gentlest slope is dominant with respect to the gas one (Azzopardi and Hills, 2003). Given the existing relationship between the size of gas slugs and their overpressure at burst (e.g. Del Bello et al., 2012), we can infer that the stronger explosive activity observed at Crater N reflects the partitioning in the uppermost conduit and preferential distribution rise of the gas-rich phase toward a steeper or wider branch of the conduit feeding this crater.

Similarly, also the differences in explosion frequency (i.e. the inverse of inter-arrival times) observed at Crater N and S are coherent with experimental observations on gas-liquid flow partitioning. In two-phase wavy-flow systems, the frequency of slug arrival is strongly affected by the inclination angle and the diameter of the pipe (Hernandez-Perez et al., 2010). Experiments performed on slug flow along inclined pipes (0-90 degrees from the horizontal) revealed that slug frequency reaches its maximum (i.e. short slugs propagating at high velocities) for inclination angles between 50 and 80 degrees (Van Hout et al., 2003; Hernandez-Perez et al., 2010) and it increases with increasing liquid flow rate (Hernandez-Perez et al., 2010). In light of this, the higher explosion frequency (i.e. shorter inter-arrival times) displayed by Crater S with respect to the other vent, is ascribed to the inclination of the branched arm with respect to the nearly vertical conduit of Crater N. Increase in the flow rate cause changes in the gas partitioning at the bifurcation, triggering a switch from puffing to Strombolian activity at Crater S (**Fig. 9**). The time variations of the explosion frequency, mostly observed at Crater S during such switches likely reflect the changes in size and release frequency of pressurized gas pockets (slugs) at depth, regulated by magma supply rate (Ripepe et al., 2002; Taddeucci et al., 2013). The hypothesis of lower gas rate, invoked at Crater S to explain such unsteady activity, is consistent with literature findings; indeed, at inclined conduits with more gentle slope the transition from bubbly to slug flow occurs at lower gas supply rate (James et al., 2004) than at vertical conduit.

Hence, the switching between Strombolian and puffing activity observed at Crater S might reflect the temporal changes at depth in flow rate, accommodated by redistribution of the gas-liquid phases at the two branches (**Fig. 9**). Similarly, periodic variations of the frequency of gas bursts/puffing have been identified at Stromboli from thermal and infrasonic evidences, and ascribed to differences in the magma supply rate (Ripepe et al., 2002).

The shift from explosive to puffing activity is marked also by a decrease of the unfiltered/filtered thermal signal ratio, indicating a lower pyroclasts to gas ratio (**Fig. 3c**). Smaller gas pockets –or slugs- hold less overpressure and are thus less efficient at fragmenting and ejecting pyroclasts (e.g. Del Bello et al., 2012). The use of such a simple thermal ratio can contribute to monitor efficiently changes in activity style during Strombolian activity and to the interpretation of the geophysical signals.

**4.2 Short-term history of the multi-vent system from integrated thermal and acoustic data**

The thermal characterization of the vent activities was fundamental for the interpretation of the infrasonic data. Because of the proximity of such vents and of the relatively distant microphones belonging to the permanent network, as well as of the weak amplitudes of most of the infrasonic events, it was not possible to precisely locate the sources and distinguish the infrasonic events generated by each vent. Hence, all the above shown infrasonic results refer to the infrasonic activity of both Crater N and S, as it was also not possible to quantify the temporal distribution of the infrasonic activity of each vent separately. Since thermal peaks and infrasonic events are different expressions of the same phenomenon (puffing/Strombolian activities; e.g. Ripepe et al., 2002), on the basis of the results obtained analysing the thermal time series, we interpreted the bimodal distribution of the infrasonic inter-arrival times by comparing it to a simulated synthetic inter-arrival times distribution.

We assumed that the dataset of infrasonic events distributes according to the same ratio between explosive events from Crater N and S observed by the thermal camera and equal to 0.8 (4,100 and 5,100 events from Crater N and S, respectively). Therefore, to reproduce the behaviour of the recorded series of 77,000 infrasonic events, we generated two synthetic series of 34,000 and 43,000 (same ratio equal to 0.8) normally-distributed inter-arrival times with average values equal to the median of the inter-arrival times distribution at Crater N and S (2.3 s and 1.6 s, respectively; **Fig. 5a,b**). We considered deviation standards equal to 25% of the average value. Next, we attributed two distinct series of synthetic normally distributed arbitrary amplitude values to both the synthetic inter-arrival time series. We assigned a central value of the synthetic amplitude series attributed to Crater N synthetic inter-arrival times series 1.5 times bigger than at Crater S. This assumption is justified by the evidence that both visual observation and thermal analyses pointed to Crater N as the strongest explosive vent. By applying first a cumulative sum separately to the two inter-arrival time series and successively merging them, we obtained a list of detection times of such synthetic events. Finally, infrasonic signals, as all the geophysical signals, show noise (even in this experiment, notably characterised by very high signal to noise ratio) preventing to detect the smallest events. Accordingly, we filtered the dataset and deleted the 10% of the detection times with the smallest amplitude values.

**Fig. 10a**, representing the histogram of the inter-arrival times of the resulting synthetic detections, shows a distribution pattern, similar to the one obtained for the observed infrasonic inter-arrival times (**Fig. 8a**), with a peak centred at 1.1-1.2 s and a slowly-decaying tail extending up to 4 s. This suggests the presence of another hidden peak with longer inter-arrival times. Since we attributed a synthetic amplitude value to each synthetic inter-arrival time, we could also simulate the inter-arrival time distribution of the detections with amplitude higher than $55^{th}$ percentile (**Fig. 10b**). Similar to the recorded infrasonic events (**Fig. 8b**), the synthetic higher amplitude detections show a unimodal distribution pattern of the inter-arrival time centred at 2.3 s.

Based on such results, the bimodal distribution of the recorded infrasonic inter-arrival times is interpreted as follows: the smaller peak centred at 2.2 s (**Fig. 8a**) is related to the Crater N infrasonic events, with larger amplitude and lower occurrence rate, while the higher peak centred at 0.7 s (**Fig. 8a**) is likely to result from the superimposition of the relatively stable Crater N infrasonic events with the smaller, more frequent and variable Crater S infrasonic events. Hence, the alternating preponderance of the shorter and longer inter-arrival times (**Fig. 8c**), highlights the temporal variability of the dominance of the infrasonic detections from Crater N or S. In particular, during the periods of puffing at Crater S (e.g. 07:15-07:47 and 10:45-11:10 on July $16^{th}$; **Fig. 3a**) the peak inter-arrival times became longer (from less than 1 s to more than 2 s; **Fig. 8c**). The small puffing infrasonic events are less detectable and the inter-arrival times distribution is dominated by the stronger and less frequent infrasonic events from Crater N. Conversely, periods of Strombolian activity at Crater S (e.g. 07:45-10:30 and 11:00-12:30 on July $16^{th}$; **Fig. 3a**) are characterised by shorter peak inter-arrival times (**Fig. 8c**), because more infrasonic events from Crater S are detected and mixed with events from Crater N. Thermal analysis, which discriminates the activity at the two vents, clearly shows how the onset of puffing activity at Crater S results in a decrease in the peak inter-arrival time (increase in the bursting rate).

Based on such evidence, we can extend the characterization of the style of activity at the two craters to the period when only Station 2 was active. The broad distribution of inter-arrival time noted during the night of July 15[th] is representative of the predominance of puffing activity at Crater S, enhancing the detections of event at Crater N, and therefore, the dominance of the peak at ca 2.2 s in **Fig. 8c**. During the same period, the lowest values of COV were observed for the most energetic infrasonic events (above the 50[th] percentile), suggesting the quasi-periodic behaviour of the source of high amplitude events, that, according the above-mentioned consideration, is likely to correspond with Crater N.

**5. Concluding remarks**

The joint thermal-infrasonic investigation of the fissure-fed multi-vent eruptive activity of July 2014 at Etna volcano has enabled us to elucidate the short-term dynamics of the plumbing system, as follows:

1. The thermal investigation of the eruptive style at the two vents active along the EF allowed us to determine the individual behaviour of each vent. In particular, Crater N featured explosions with higher median amplitude and longer recurrence time than Crater S. Different switches from puffing to Strombolian activity (and vice-versa) at Crater S were tracked via thermal analysis and confirmed by visual observations. Overall a bimodal distribution of thermal amplitude characterizes this vent. In contrast, Crater N featured a remarkably stable periodicity.

2. The temporal evolution of inter-arrival times and amplitudes at the two eruptive vents are significantly correlated, i.e. the dynamics at the two vents are not independent. Based on this evidence and on the eruptive style of each vent (recurrence time, thermal amplitude, etc.) we inferred the feeding structure of the multi-vent system to consist of a branched conduit. The longer recurrence time and higher amplitude of explosive activity observed at Crater N, both in thermal analyses and visual observation, probably resulted from preferential uprise of the gas phase through the steeper conduit. We inferred that the switches from puffing to Strombolian activity accommodated an increase in the supply rate, in response to the redistribution of gas and liquid phases at the conduit split.

    3.    Comparing thermal data with infrasound observations, we were able to interpret the distribution of inter-arrival times observed in the acoustic signals as an overlap between Crater N and Crater S inter-arrival time series. Based on comparison of the acoustic data with synthetic series, we additionally inferred the predominance of puffing activity at Crater S during the night of July $15^{th}$.

*Acknowledgements*

The data used in this work were collected during experiments performed by the authors and they may be available upon request to the corresponding author. This project was supported by the EU-funded FP7 project MEDiterranean SUpersite Volcanoes (MED-SUV), grant agreement 308665. L.S. wish to thank the ERC Consolidator "CHRONOS" project (Grant No. 612776). L.S. and A.C. thank the project AEOLUS funded by the Fondo di Ricerca di Base of Department of Physics and Geology, University of Perugia. U.K. additionally acknowledges the support by the EU-funded Initial Training Network 'VERTIGO', grant agreement 607905. DBD acknowledges the support of the ERC Advanced Grant EVOKES (247076). The authors are grateful to the editors G. Lube and J.D.L. White and to an anonymous reviewer for the detailed comments and fruitful suggestions. We wish to thank L. Scuderi for the precious help provided during fieldworks.

**Figure captions**

**Fig. 1. (a)** Aerial view, acquired by a DJI Phantom 2 drone looking north-westwards towards the North East Crater (summit not visible), of the two active vents (Crater N and Crater S) and the geometry of the temporary setup on July $15^{th}$-$16^{th}$, 2014 (the distance between the two stations is a few meters). **(b)** Digital elevation model of Mt. Etna. The area shown in (a) is marked by a black solid rectangle. ECNE seismo-acoustic station, belonging to the INGV-OE permanent network, is represented by a black triangle. The relative position of summit craters (VOR: Voragine; BN: Bocca Nuova; NEC: North-East Crater; SEC South-East Crater; NSEC: New South-East Crater) is shown in inset **(c)**

**Fig. 2. (a)** Time line of the eruptive activity at Etna from July $1^{st}$ to September $1^{st}$, 2014. During this time, all the summit craters with the exception of NSEC showed only various extents of outgassing. **(b)** RMS (Root Mean Square) amplitude of seismic signal recorded at the vertical component of ECNE station. The deployment site for the temporary setup is marked by the green rectangle. The red arrow represents the end of the EF eruption. **(c)** Visible and **(d)** thermal images of the two active vents. The white dotted rectangles in (d) indicate the control areas used for thermal analysis (15.9 and 11.6 meters wide over the N and S vent, respectively)

**Fig. 3. (a)** Time line of the eruptive activity at the two vents. **(b)** Thermal amplitude of the explosive events detected at each vent. **(c)** Ratio between non-filtered and low-pass filtered thermal amplitudes at each vent. The ratio increases with increasing pyroclast-gas ratio. **(d)** Inter-arrival time between pairs of consecutive events measured at the two vents. Note that discontinuous recording resulted in gaps in the time series. Numbers in the grey line in b) indicate the focal length (in mm) of the lens used for thermal acquisition .

**Fig. 4.** Example of thermal analysis of **a)** Crater N and **b)** Crater S calculated by integrating the amplitude of the thermal images in the two control-areas, and correcting for the background.

The dots represent events detected by applying a static threshold at zero (black dotted line). Zero time is July 16th, 07:28:57.000.

**Fig. 5.** Histograms of the inter-arrival times and amplitudes of triggered thermal events at Crater N (**a,c**, respectively) and Crater S (**b,d**, respectively).

**Fig. 6. (a)** Temporal variation of the inter-arrival times of the explosions detected from the thermal signals at Crater N (red) and S (blue). **(b)** Randomised cross correlation analysis between inter-arrival times of the explosions detected on the thermal signals at Crater N and S. Red dot: the cross correlation value between inter-arrival times series; blue dots: cross correlation values between inter-arrival times at Crater S and 10,000 randomized versions of the inter-arrival time series at Crater N; black dashed lines: the $\hat{c} \pm 3\sigma c$ band boundaries. **(c)** Variation in time of the amplitudes of the explosions detected on the thermal signals at Crater N (red) and S (blue). **(d)** Randomised cross correlation analysis between amplitudes of the explosions detected on the thermal signals at Crater N and S. Other symbols as in (**b**) but for amplitudes instead of inter-arrival times series

**Fig. 7. (a)** Infrasonic signal recorded during 18:27:19-18:27:49 on July 15th, 2014. **(b)** Infrasonic signal recorded from ~12:00 on July 15th to ~13:00 on July 16th, 2014 and **(c)** corresponding spectrogram. The red rectangle in (b) marks the 30-sec-long window shown in (a)

**Fig. 8. (a)** Histogram showing the total number of infrasound event pairs versus the inter-arrival time values for both craters. **(b)** Histogram showing the total number of infrasound event pairs, with peak-to-peak amplitude higher than 50th percentile, versus the inter-arrival time values for both craters. **(c)** contour plot showing time in the x-axis, inter-arrival times in the y-axis and the number of infrasound event pairs with the color scale. The grey box indicates the time interval when Station 1 was running. **(d,e)** Examples of 10-sec-long windows of infrasonic signal recorded by the temporary station. **(f)** Variation in time of Coefficient Of Variation (COV) values calculated

on the inter-arrival times between infrasonic events. **(g)** Variation in time of COV values calculated on the inter-arrival times between infrasonic events, with peak-to-peak amplitude higher than the $50^{th}$ percentile. In (a) and (b) the red arrows mark the position of the peaks in the histograms

**Figure 9.** Sketch of the shallow conduit of the two-vent system during phases of **(a)** puffing and **(b)** Strombolian activity at Crater S

**Figure 10. (a)** Histogram showing the distribution of the synthetic infrasound event inter-arrival time values, with amplitudes higher than $10^{th}$ percentile. **(b)** Histogram showing the distribution of the synthetic infrasound event inter-arrival time values, with amplitudes higher than $55^{th}$ percentile

**Figure 1**

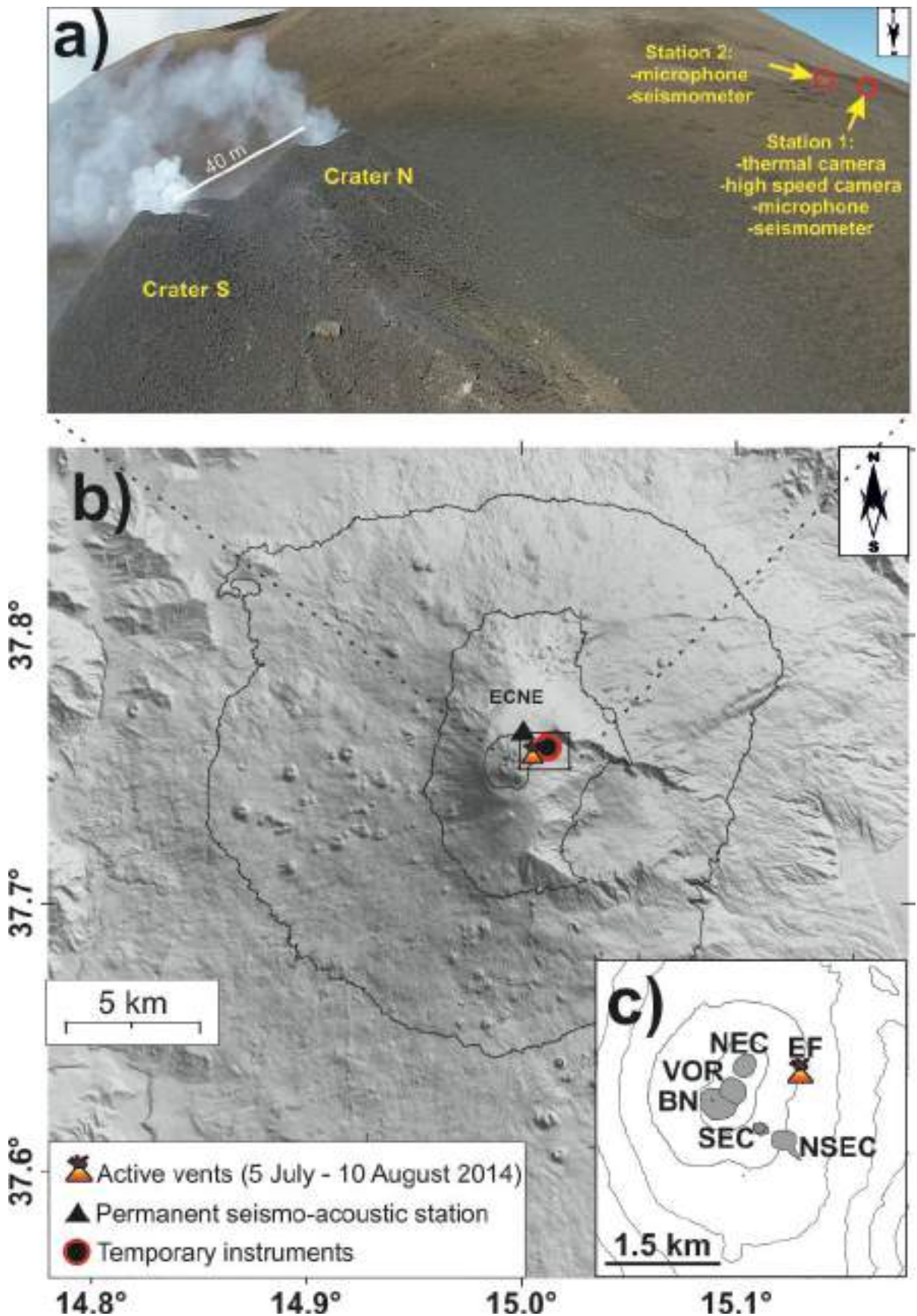

**Figure 2**

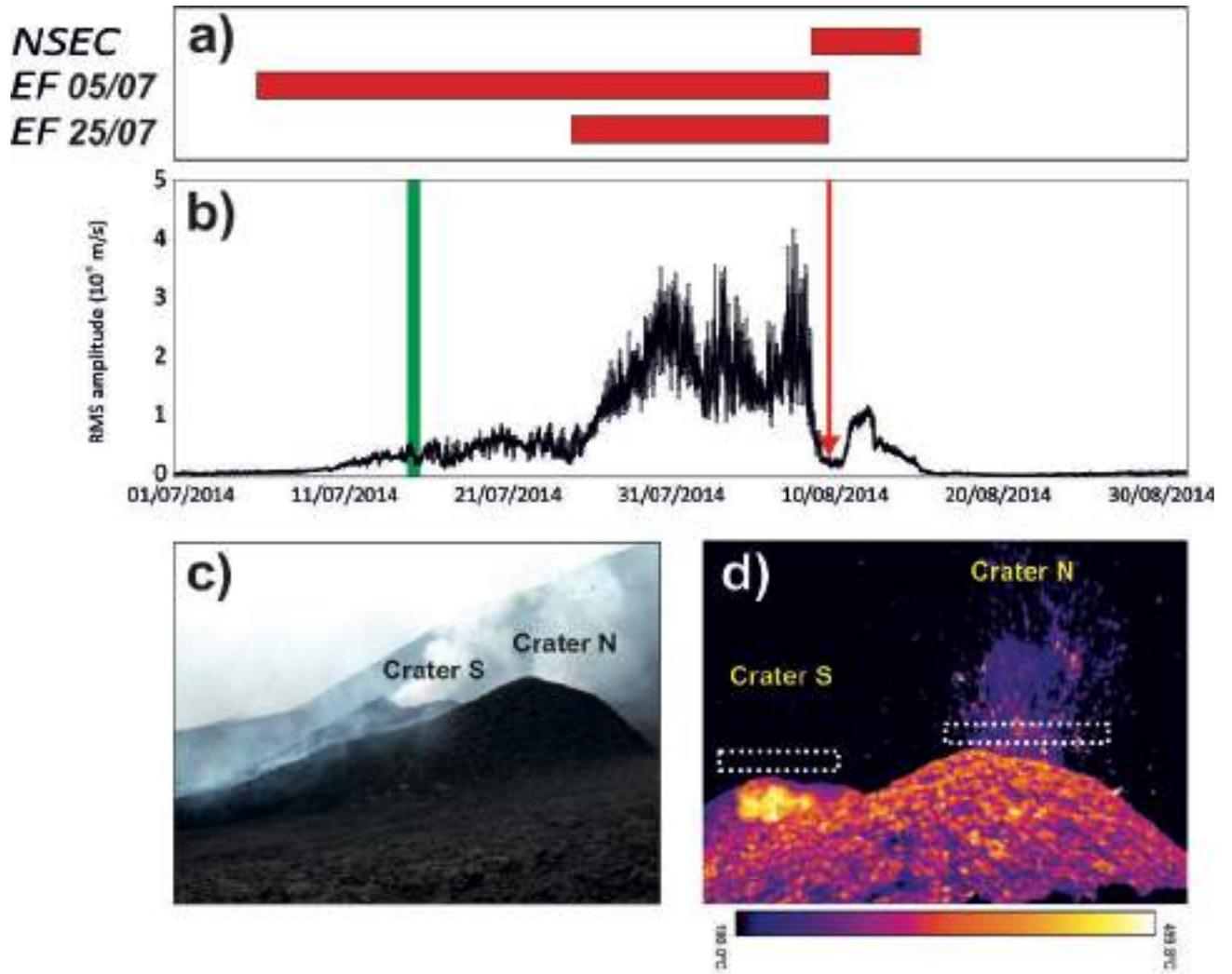

**Figure 3**

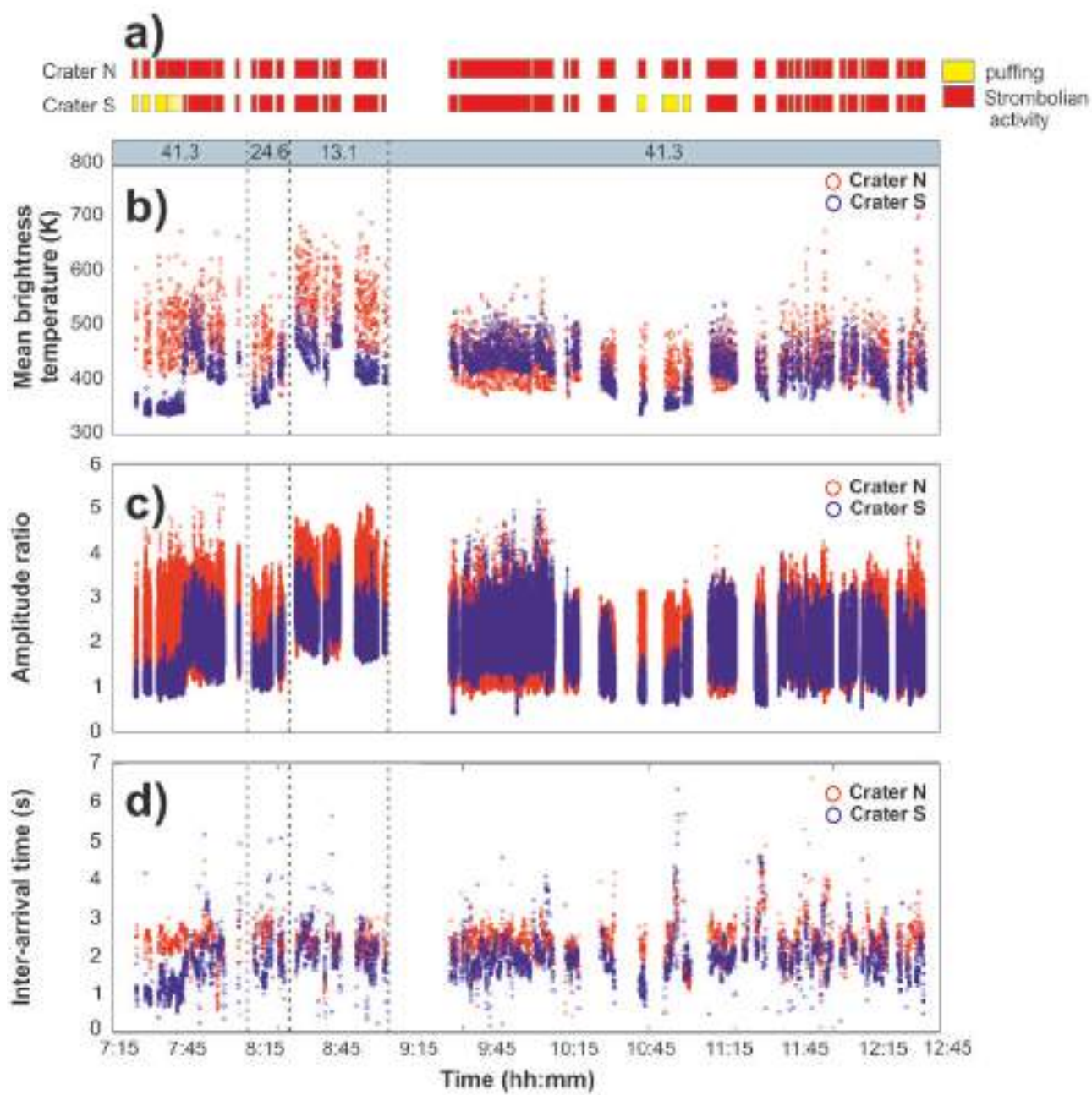

**Figure 4**

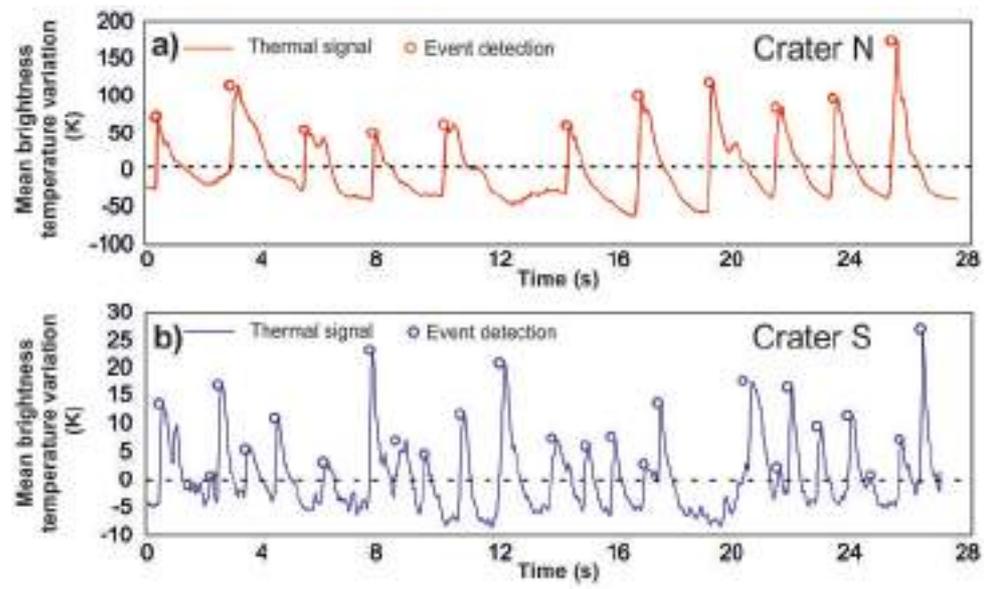

**Figure 5**

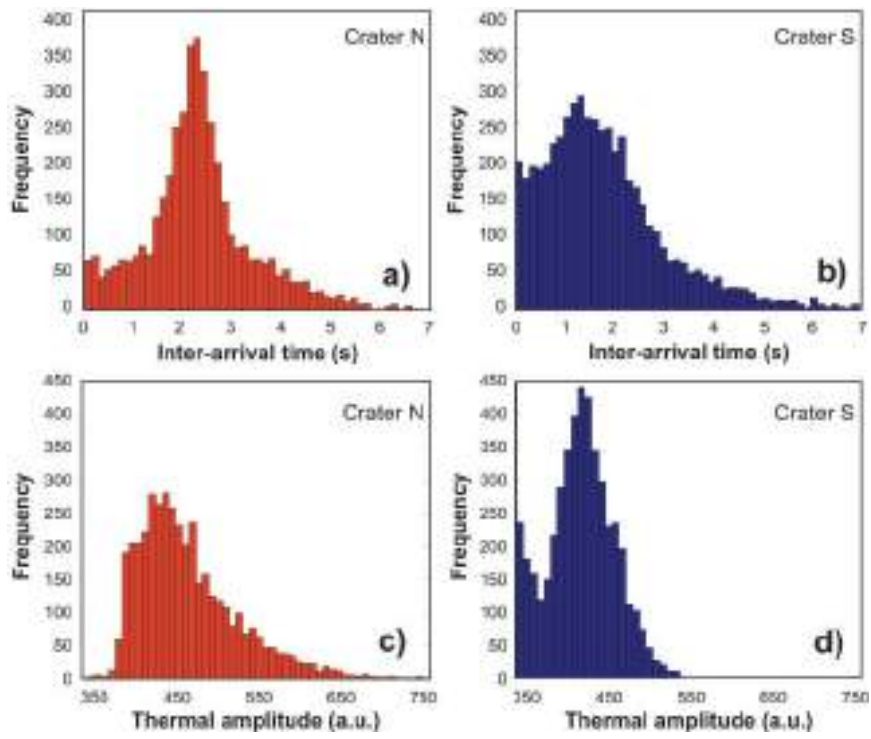

**Figure 6**

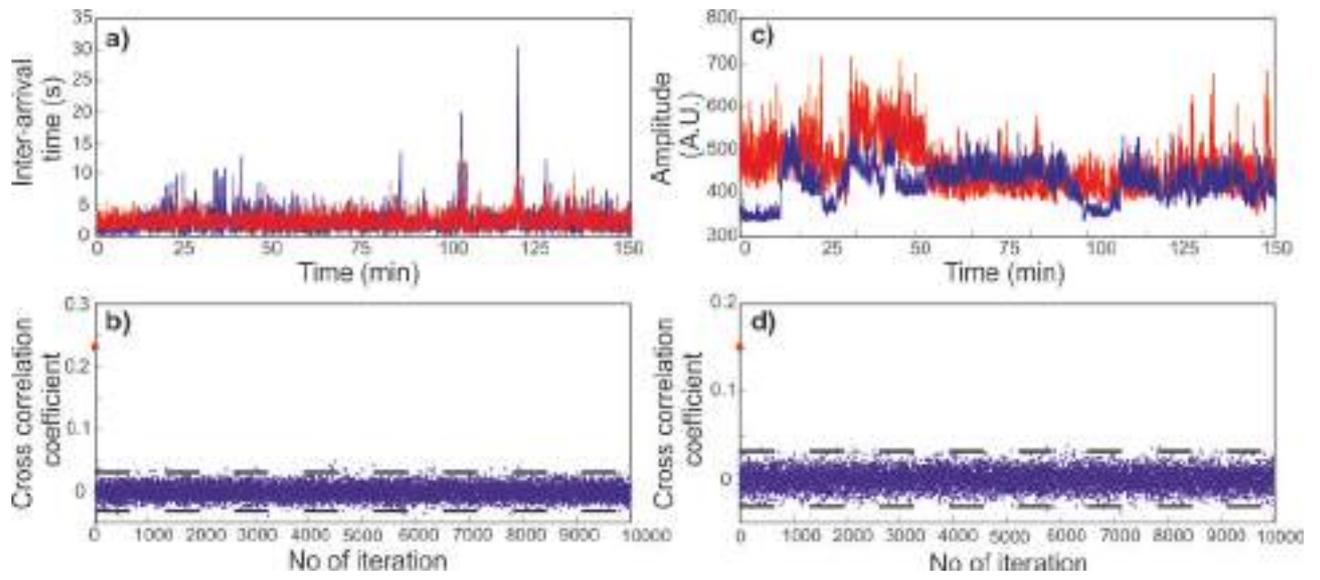

**Figure 7**

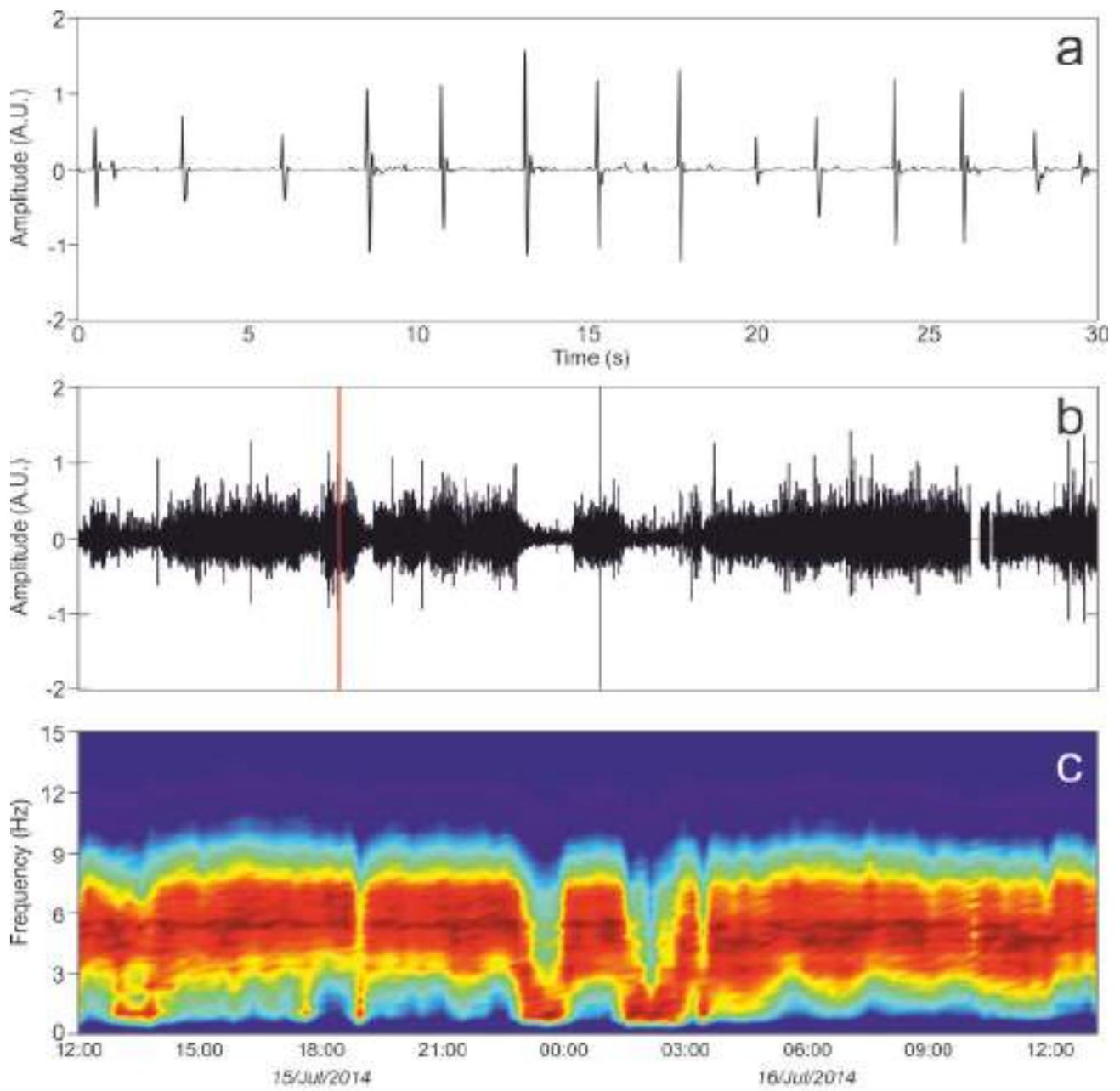

**Figure 8**

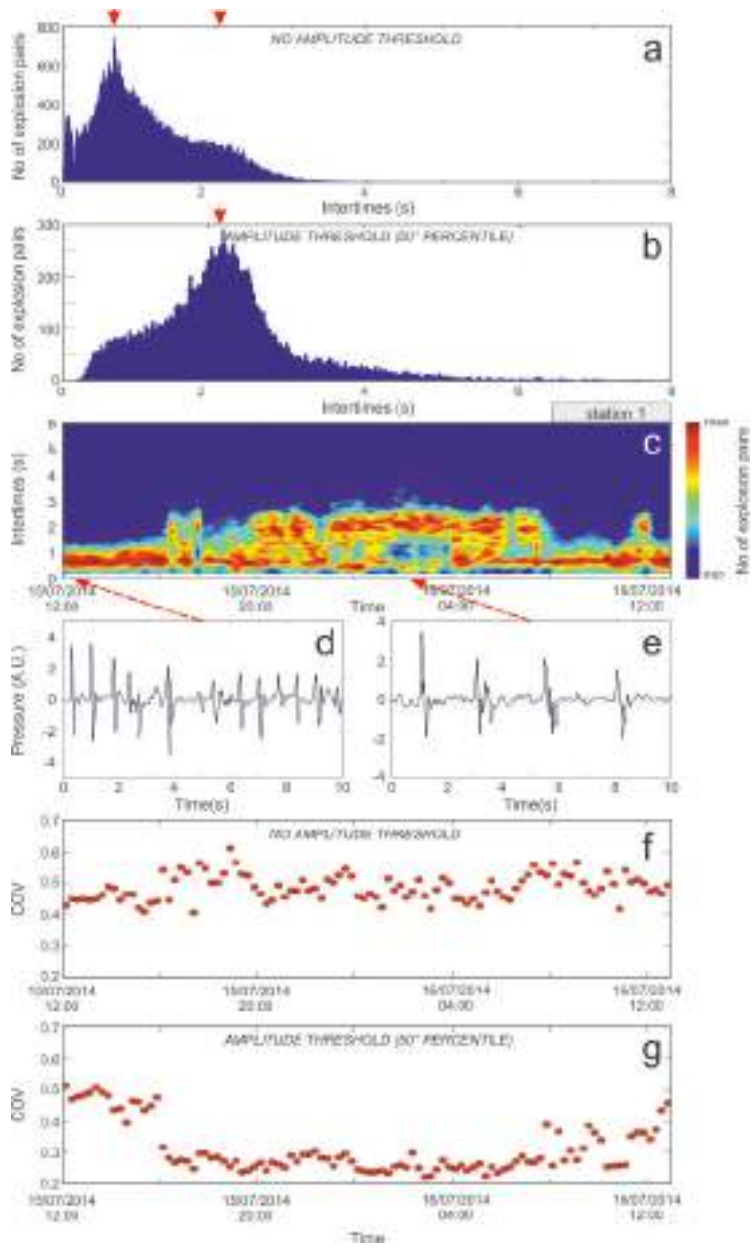

**Figure 9**

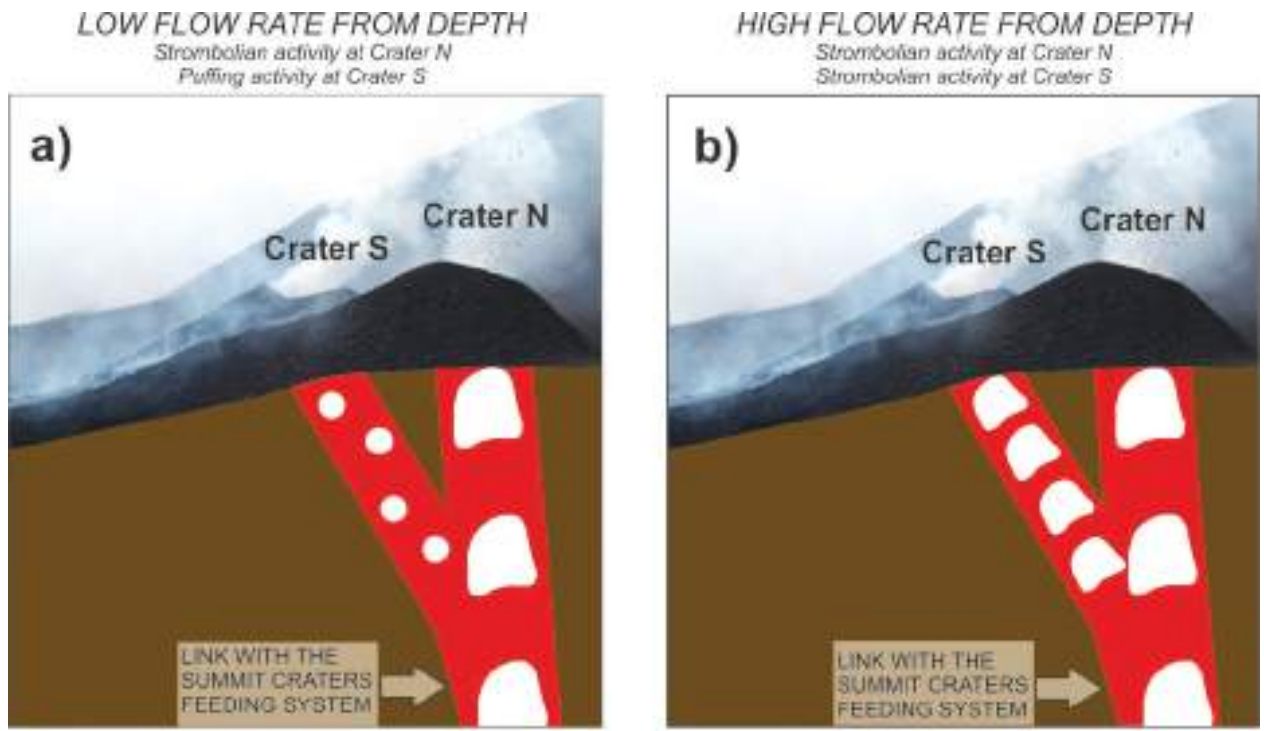

**Figure 10**

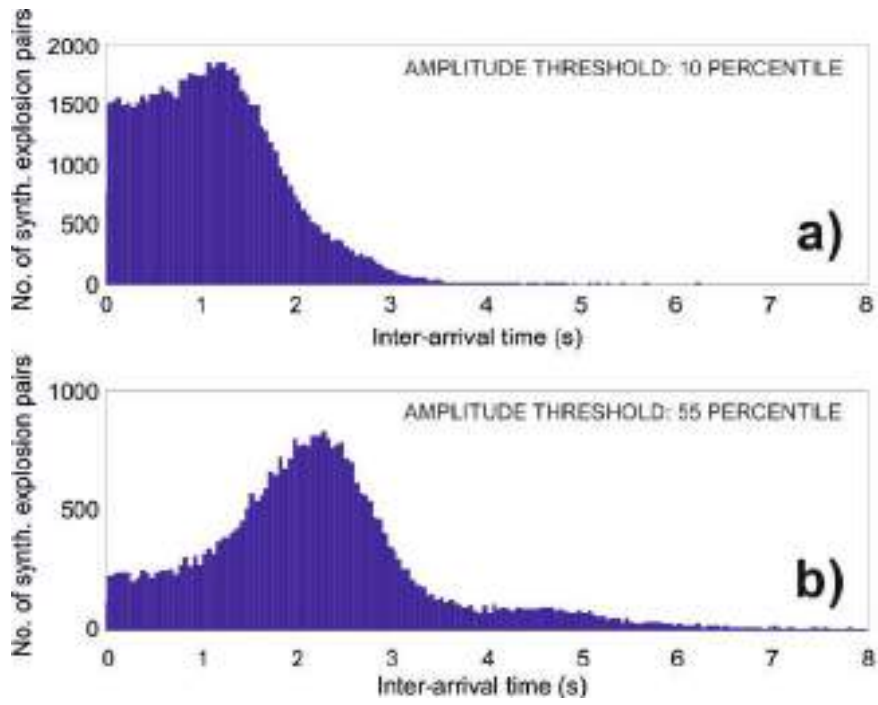